\documentclass{article}
\usepackage[colorlinks, pdfborder={0 0 0}, plainpages=false]{hyperref}
\usepackage{graphicx}
\usepackage{grffile}
\usepackage{subfigure}
\usepackage{xspace}
\usepackage[usenames,dvipsnames]{color}
\usepackage{amssymb}
\usepackage[normalem]{ulem} 
\usepackage{letltxmacro}
\usepackage{amsfonts}
\usepackage{amsmath}
\usepackage{amssymb}
\usepackage{bm}
\usepackage{dcolumn}
\usepackage{epsfig}
\usepackage{graphicx}
\usepackage{graphics}
\usepackage[latin1]{inputenc}
\usepackage{latexsym}
\usepackage{rotating}
\usepackage{hyperref}
\usepackage[usenames]{color}
\usepackage{float}
\usepackage{xspace} 
\usepackage{mathrsfs}
\usepackage{subfigure}
\usepackage{multirow}

\usepackage[T1]{fontenc}    
\usepackage{hyperref}       
\usepackage{url}            
\usepackage{booktabs}       
\usepackage{amsfonts}       
\usepackage{nicefrac}       
\usepackage{microtype}      
\usepackage[nonatbib,final]{nips_dlps_2017}

\def\Msun{M_\odot}

\def\lesssim{\mathrel{\hbox{\rlap{\hbox{\lower4pt\hbox{$\sim$}}}\hbox{$<$}}}}
\def\gtrsim{\mathrel{\hbox{\rlap{\hbox{\lower4pt\hbox{$\sim$}}}\hbox{$>$}}}}
\def\alt{\mathrel{\hbox{\rlap{\hbox{\lower4pt\hbox{$\sim$}}}\hbox{$<$}}}}
\def\agt{\mathrel{\hbox{\rlap{\hbox{\lower4pt\hbox{$\sim$}}}\hbox{$>$}}}}

\def\gta{\ifmmode {\mathbin{\lower 3pt\hbox   
			{$\,\rlap{\raise 5pt\hbox{$\char'076$}}\mathchar"7218\,$}}}
	\else {${\mathbin{\lower 3pt\hbox
				{$\rlap{\raise 5pt\hbox{$\char'076$}}\mathchar"7218\,$}}}
		$}\fi}
\def\lta{\ifmmode {\,\mathbin{\lower 3pt\hbox   
			{$\,\rlap{\raise 5pt\hbox{$\char'074$}}\mathchar"7218\,$}}}
	\else {${\mathbin{\lower 3pt\hbox
				{$\rlap{\raise 5pt\hbox{$\char'074$}}\mathchar"7218\,$}}}
		$}\fi}

\newcommand{\beq}{\begin{equation}}
\newcommand{\eeq}{\end{equation}}
\newcommand{\bea}{\begin{eqnarray}}
\newcommand{\eea}{\end{eqnarray}}

\title{Deep Learning for Real-time Gravitational Wave Detection and Parameter Estimation with LIGO Data}

\author{
	Daniel George \\
	NCSA and Department of Astronomy \\
	University of Illinois at Urbana-Champaign \\
	\texttt{dgeorge5@illinois.edu} \\
	\And
	E.~A. Huerta \\
	NCSA \\
	University of Illinois at Urbana-Champaign \\
	\texttt{elihu@illinois.edu} 
}
\begin{document}

	\maketitle

	\begin{abstract}
		
		The recent Nobel-prize-winning detections of gravitational waves from merging black holes and the subsequent detection of the collision of two neutron stars in coincidence with electromagnetic observations have inaugurated a new era of multimessenger astrophysics. To enhance the scope of this emergent science, we proposed the use of deep convolutional neural networks for the detection and characterization of gravitational wave signals in real-time. This method, \texttt{Deep Filtering}, was initially demonstrated using simulated LIGO noise. In this article, we present the extension of \texttt{Deep Filtering} using real data from the first observing run of LIGO, for both \textit{detection} and \textit{parameter estimation} of gravitational waves from binary black hole mergers with continuous data streams from multiple LIGO detectors. We show for the first time that machine learning can detect and estimate the true parameters of a real GW event observed by LIGO. Our comparisons show that \texttt{Deep Filtering} is far more computationally efficient than matched-filtering, while retaining similar sensitivity and lower errors, allowing real-time processing of weak time-series signals in non-stationary non-Gaussian noise, with minimal resources, and also enables the detection of new classes of gravitational wave sources that may go unnoticed with existing detection algorithms. This approach is uniquely suited to enable coincident detection campaigns of gravitational waves and their multimessenger counterparts in real-time.

	\end{abstract}

	
	\section{Introduction} 
	\label{intro}
	
	The detection of gravitational waves (GWs) from the merger of binary black holes (BBHs), and the subsequent detections of neutron star mergers, by LIGO~\cite{LSC:2015} has set in motion a scientific revolution~\cite{DI:2016,secondBBH:2016,bbhswithligo:2016,thirddetection,2017arXiv170909660T,BNSdet:2017}. However, matched-filtering, the most sensitive search algorithm used by LIGO, targets only a 3D parameter space (spin-aligned binaries on quasi-circular orbits)~\cite{2013PhRvD..87j4028G,Carl:2016arXiv,CR:2015PRL}---a subset of the over 8D parameter space of GW signals~\cite{Anto:2015arXiv,Samsing:2014,Huerta:2017a,Lehner:2014a}. Extending these searches to target eccentric~\cite{Sergey:2016,Huerta:2017a,Huerta:2014,Huerta:2013a} or spin-precessing BBHs is computationally infeasible~\cite{2016PhRvD..94b4012H}. 
	
	To address these limitations, a Deep Learning~\cite{DL-Nature} technique called \texttt{Deep Filtering} was introduced by George and Huerta~\cite{DeepFiltering}. This method employs a system of two deep convolution neural networks (CNNs~\cite{lecun98-cnn,DL-Nature}) that directly take time-series inputs for classification and regression. In their first article~\cite{DeepFiltering}, they showed that CNNs can outperform traditional machine learning methods, reaching sensitivities comparable to matched-filtering for directly processing highly noisy time-series data streams to detect weak GW signals and estimate the parameters of their source in real-time using GW signals injected into simulated LIGO noise. 
	
	In this article, we extend \texttt{Deep Filtering} to GW signals in real LIGO noise~\cite{DNNRealNoise}. We prove that CNNs can be used for both signal detection and multiple-parameter estimation directly from extremely weak time-series signals embedded in highly non-Gaussian non-stationary noise from LIGO. Our results indicate that CNNs achieve similar performance to matched-filtering methods, while being many orders of magnitude faster and far more resilient to transient anomalous noise artifacts such as glitches in LIGO. This article shows for the \textit{first} time that machine learning can successfully recover \textit{true} GW signals observed by LIGO. Furthermore, we show that after a single training process, \texttt{Deep Filtering} can generalize to different LIGO noise with new Power Spectral Densities (PSDs), without re-training.
	
	We also show that \texttt{Deep Filtering} can interpolate between templates, \textit{generalize} to new classes of signals beyond the training data, and also detect signals and measure their parameters, even when contaminated by glitches. We show that it is more robust than matched-filtering especially in the presence of anomalies, which indicates its applicability for glitch classification and clustering efforts critical for GW detector characterization. Since all the intensive computation is diverted to the one-time training stage, template banks of any size may be used for training after which data streams can be analyzed in real-time with a single CPU or GPU. \texttt{Deep Filtering} can be potentially used to rapidly narrow down the parameter space of GW detections, which can then be quickly followed up with existing pipelines using a few templates around the predicted parameters to measure the significance of the events and obtain more informative parameter estimates, thus accelerating existing GW analysis, with minimal computational resources, across the full range of GW signals. 
	
	
	\section{Methods}
	\label{meth}
	
	\texttt{Deep Filtering} consists of two steps: a classifier CNN is first applied to the data stream via a sliding window of 1s width and a step size of 0.2s. The classifier returns probability of the presence of a signal. If the classifier detects a signal, a predictor CNN is applied to the input to determine the parameters such as masses. In a multi-detector scenario, the CNNs may be applied separately to each data stream and coincidence of detections with similar parameters can be enforced, which can then be verified quickly by matched-filtering with the predicted templates. 
	
\begin{figure*}
		\centerline{
		\hspace{-.1in}\includegraphics[width=.484\textwidth]{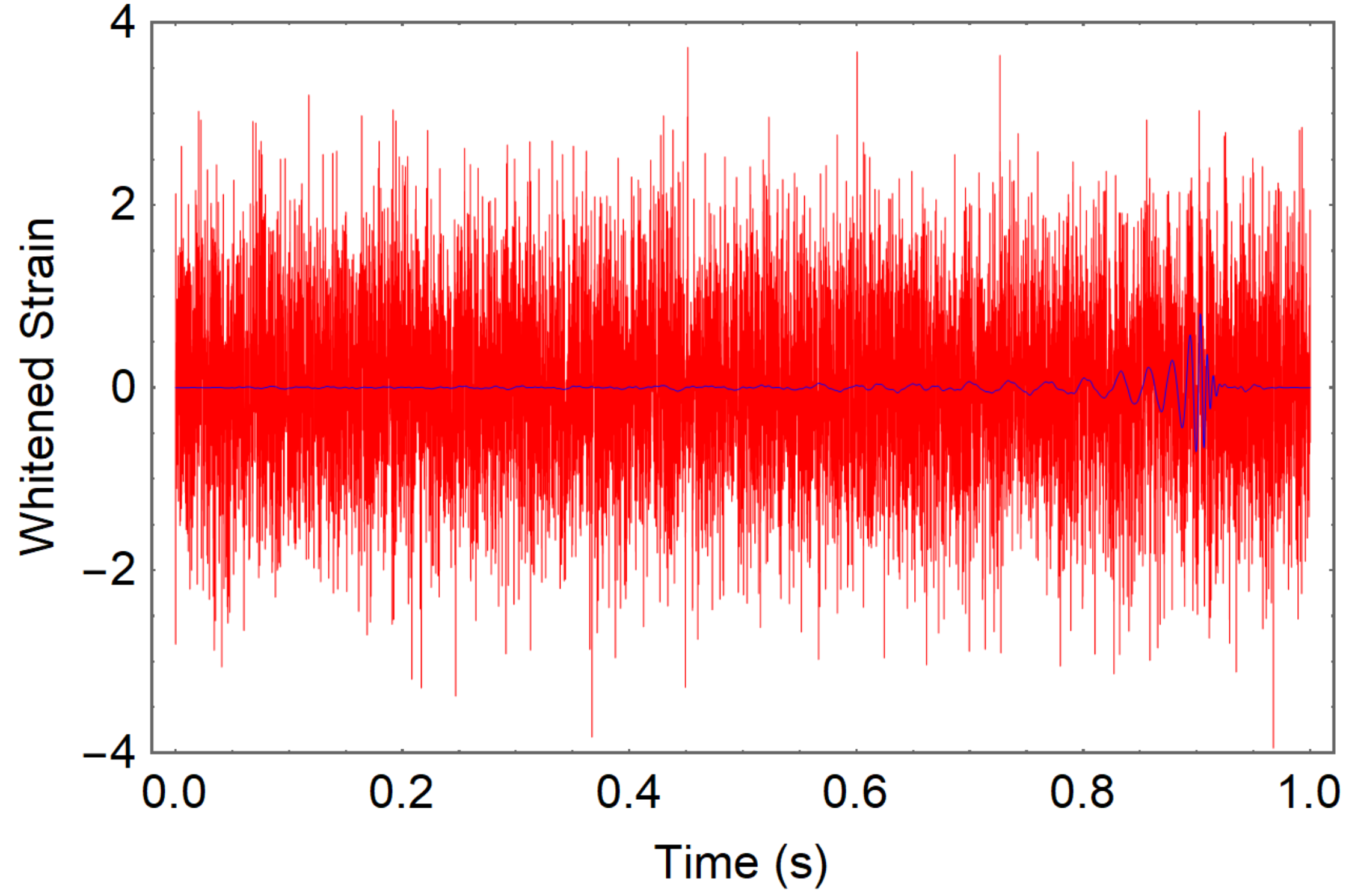}\hspace{.2in}
			\includegraphics[width=0.484\textwidth]{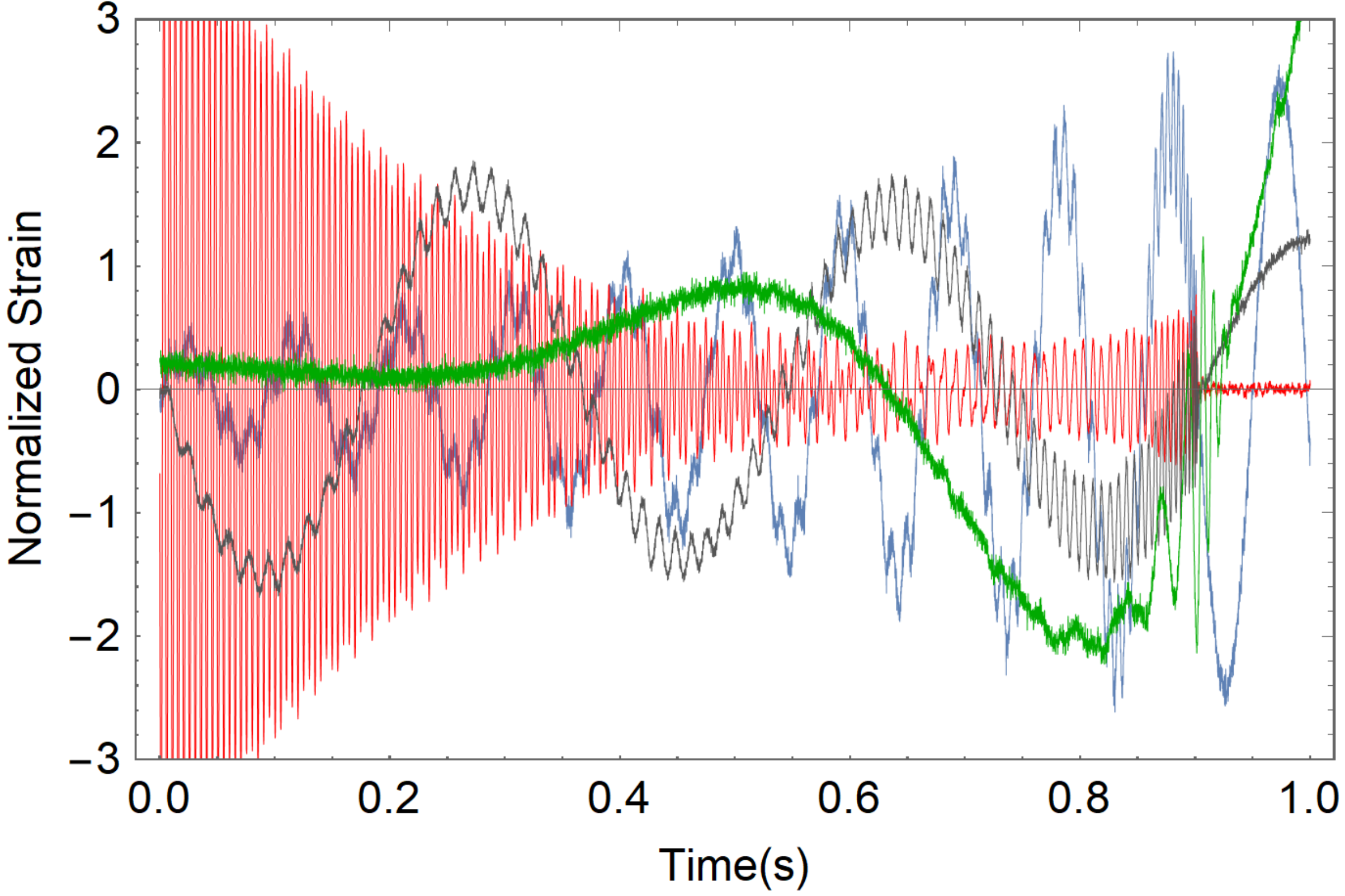}
		}
		\caption{The red time-series on the \textbf{left} is an input to our CNNs. It contains a BBH GW signal (blue) from our test set which was superimposed in real LIGO noise. For this injection, the optimal matched-filter SNR = 7.5 (peak power of this signal is 0.65 times the power of background noise). The presence of this signal was detected directly from the (red) time-series input with over 99\% sensitivity and the source's parameters were estimated with a mean relative error less than 10\%. On the \textbf{right}, we show some GW signals that happen to occur in coincidence with glitches. \texttt{Deep Filtering} can accurately identify these signals contaminated by glitches, unlike existing methods.}
		\label{fig:signal}
	\end{figure*}

	The datasets of waveform templates used to train and test our CNNs were obtained with the EOB code~\cite{Tara:2014}. Our training set contained about 2500 templates, with BBHs component masses chosen in the range $5\Msun$ to $75\Msun$ in steps of $1\Msun$. The component masses in the test set were separated from the training set by $0.5\Msun$ each. We obtained real LIGO data from the \href{https://losc.ligo.org/about/}{LIGO Open Science Center} (LOSC), 4096s from each detector around each of the first 3 GW events: GW150914, LVT151012, and GW151226. Noise from GW151226 and LVT151012 was used for training/validation and noise from GW150914 was used for testing. This tests the ability to generalize to different noise distributions since the PSD of LIGO varies widely with time. The inputs were whitened with the average PSD of the real noise measured at that time-period. The testing results were measured using injections of test set GW templates into pure LIGO noise not used in training (see Fig.~\ref{fig:signal}).
	
	We used similar hyperparameters to the original CNNs in~\cite{DeepFiltering}. We have 4 convolution layers with the filter sizes to 64, 128, 256, and 512 respectively and 2 fully connected layers with 128 and 64 neurons. The ReLU activation was used throughout. We used kernels of 16, 16, 16, and 32 for the convolutions and 4 for all the (max) pooling layers. Stride was 1 for all the convolutions and 4 for all the pooling layers. We observed that dilations~\cite{dilatedCNN} of 1, 2, 2, and 2 in the convolutions improved the performance. A softmax layer is used in the classifier to obtain probabilities as the outputs.
	
	For the training process, we used the curriculum learning strategy in~\cite{DeepFiltering} by starting off training GW injections at high matched filter SNR~\cite{saton} ($\ge100$) and then gradually increasing the noise during the training until a final SNR distributed in the range 4 to 15. This ensured that the performance of prediction can be quickly maximized for low SNR, while retaining performance at high SNR. We first trained the predictor on the datasets labeled with the BBH masses and then copied this network and trained it on datasets having 90\% pure random noise inputs, after adding a softmax layer. This transfer learning procedure decreases the training time for the classifier and improves its sensitivity. 
	
	
	\section{Results} 
	\label{result}

	\begin{figure*}
		\centerline{
		\hspace{-.05in}\includegraphics[width=.485\textwidth]{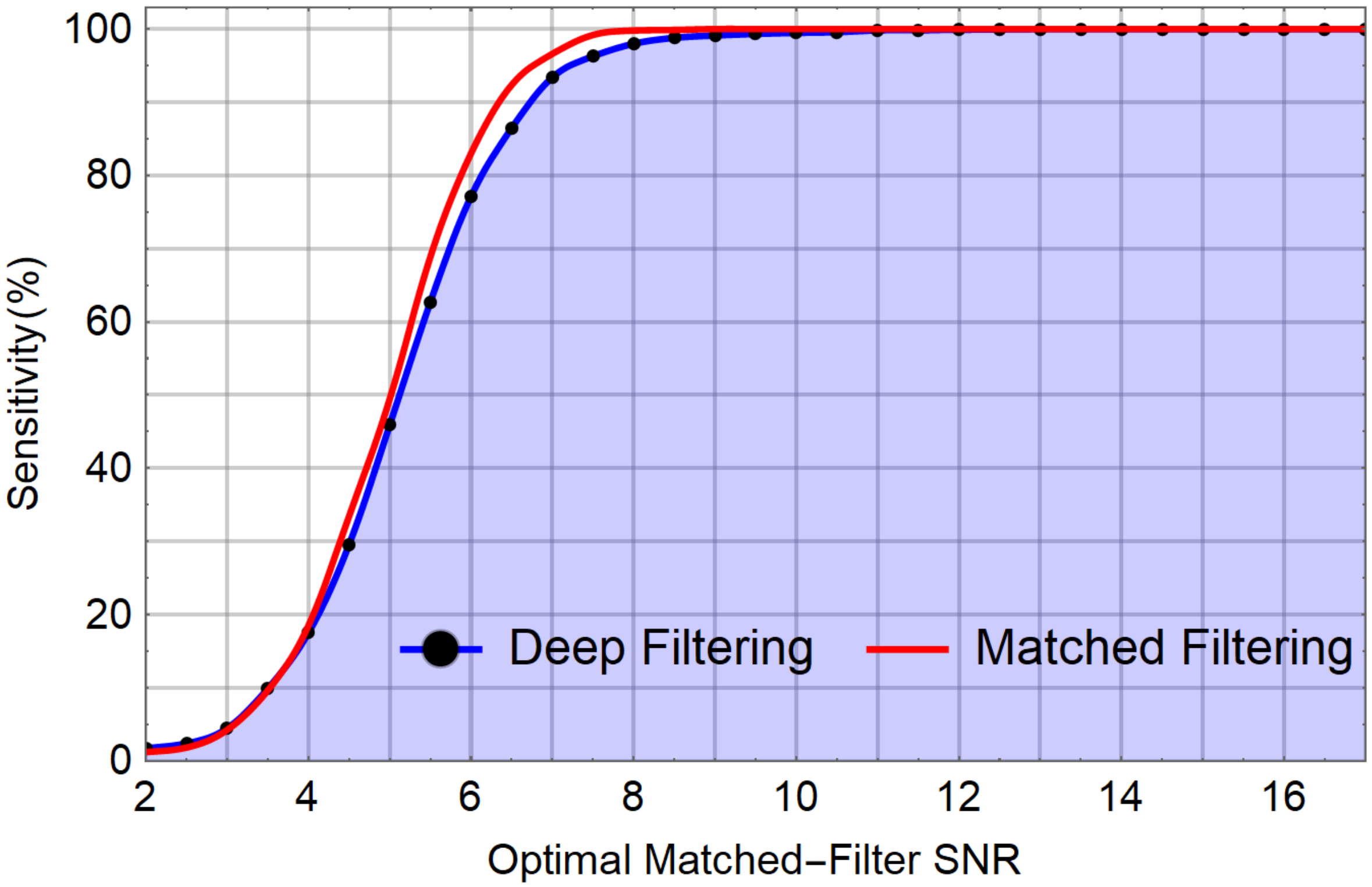}\hspace{.15in}
			\includegraphics[width=.49\textwidth]{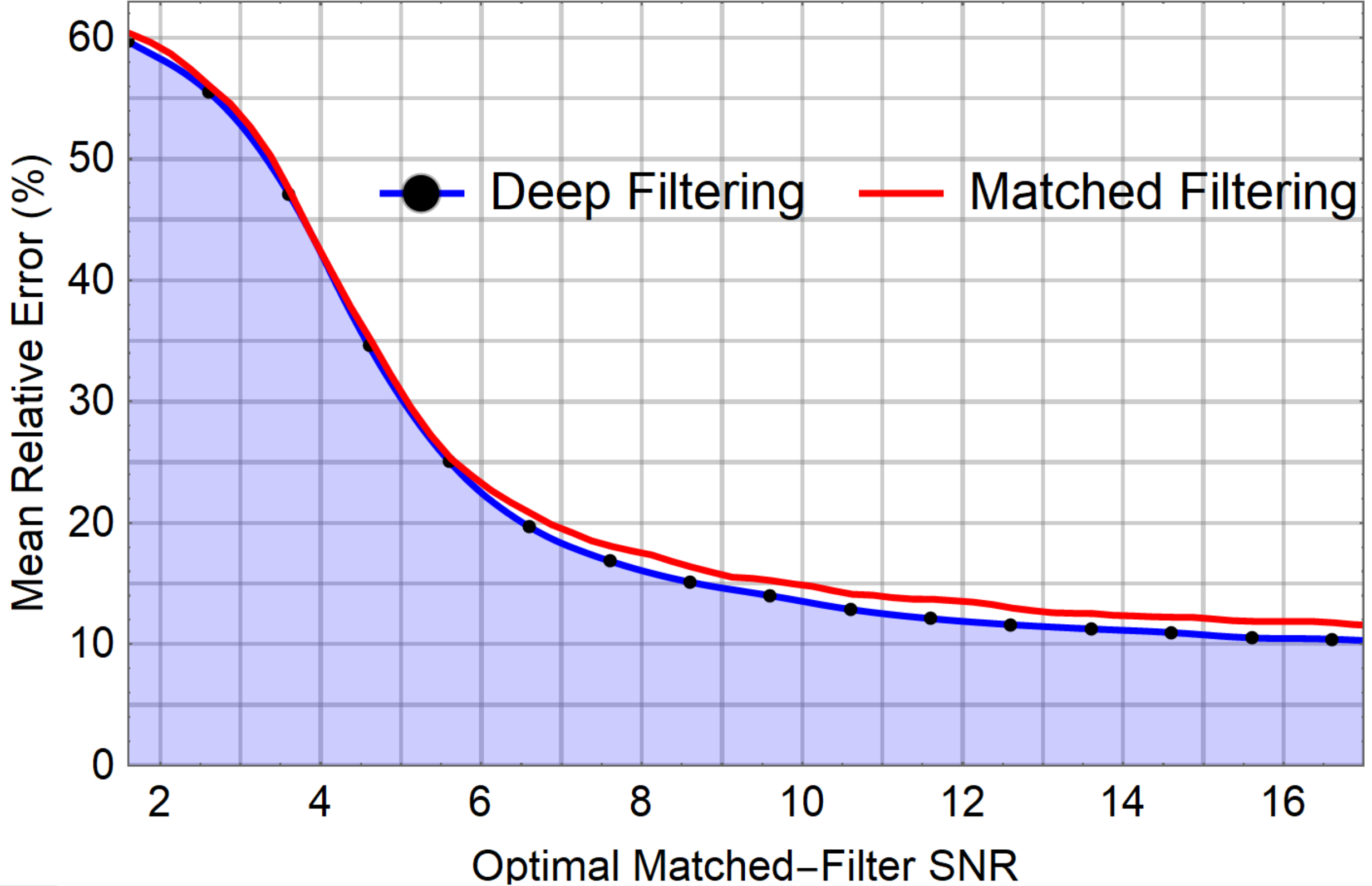}
		}
		\caption{
			The \textbf{left} panel shows the sensitivity of the classifier for detecting GW signals in real LIGO noise compared to matched filtering with the same template bank used for training. The SNR is on average $10\pm1.5$ times the ratio of the amplitude of the signal to the standard deviation of noise. This implies that signals significantly weaker than the background noise can be detected. On the \textbf{right} we compare the relative error in estimating the masses with our predictor and with matched-filtering. We can see that the predictor is able to interpolate to test set signals with intermediate parameter values. While matched-filtering has the advantage of being optimized with the PSD of the LIGO noise in the test set, \texttt{Deep Filtering} was only trained on noise from other events, therefore our results demonstrate the ability of the CNNs to automatically generalize to non-stationary LIGO noise having different PSDs without retraining.
		}
		\label{fig:Sensitivity}
	\end{figure*}
	
	The sensitivity of the classifier vs. SNR is shown in Fig.~\ref{fig:Sensitivity}. We achieved 100\% sensitivity for SNR > 10. The false alarm rate was tuned to be less than 1\%, i.e., 1 per 100 seconds of noise in our test set was classified as signals. Assuming independent noise from multiple detectors, this implies the 2-detector false alarm rate would be less than 0.01\%, when Deep Filtering is applied independently to each detector and coincidence is enforced. The false alarm rate can be further decreased by running matched-filtering pipelines with our predicted templates can eliminate these false alarms.
	
	Our predictor was able to successfully measure the component masses given noisy GW signals, that were not used for training as shown in Fig.~\ref{fig:Sensitivity}. We observed that the errors follow a Gaussian distribution for SNR greater than 10. For high SNR, we our predictor achieved mean relative error less than 10\%, whereas matched-filtering with the same template bank always has error greater than 10\%. This implies that Deep Filtering is capable of interpolating between points in the training data.
	
	Although, we trained only on simulated GW injections, we applied \texttt{Deep Filtering} to the 4096s data stream containing a true GW signal, GW150914, using a sliding window of 1s width with offsets of 0.2s through the data around each event from each detector. This signal was correctly identified by the classifier at the true position in time and each of the predicted component masses were within the published error bars~\cite{DI:2016}. There were zero false alarms after enforcing the constraint that the detection should be made simultaneously in the inputs from multiple detectors. A demo showing the application of \texttt{Deep Filtering} to GW150914 can be found here: \href{http://tiny.cc/CNN}{tiny.cc/CNN}.
	
	Furthermore, we tested the resilience of \texttt{Deep Filtering} to transient disturbances, with a simulated set of sine-Gaussian glitches, which cover a broad range of transient morphologies, following~\cite{jade1:2016}. We ensured that a different set of frequencies, amplitudes, peak positions, etc., were used for training and testing. We then injected these glitches into the training process and found that the classifier network was easily able to distinguish new glitches in the test set from true signals, with a false alarm rate less than 1\%. When we applied the standard matched-filtering algorithm to the same test set of glitches, approximately 30\% of glitches were classified as signals due to their high SNR. 	
		
	 We then tested the performance of \texttt{Deep Filtering}, when a signal is superimposed with both a glitch and real LIGO noise. We injected glitches from the training set into the training process and measured the sensitivity of the classifier on the test set signals superimposed with glitches sampled from the test set of glitches. Over 80\% of the signals with SNR of 10 were detected even after they were superimposed with glitches. These results are very promising, since we may be able to detect GW signals that occur during periods of bad data quality in the detectors using \texttt{Deep Filtering}, whereas currently such periods are vetoed and left out of the analysis by LIGO pipelines.
	
	Another experiment that we tried was to inject waveforms obtained from simulations of eccentric BBH systems on Blue Waters, with eccentricities between 0.1 and 0.2 when entering the LIGO band, that we performed using the open-source Einstein Toolkit~\cite{ETL:2012CQGra,ENIGMA}, as well as waveforms from spin-precessing binaries from the public SXS catalog~\cite{chu:2016CQG}. We found that these signals were detected with the same sensitivity as the original test set of quasi-circular BBH waveforms by our classifier, achieving 100\% sensitivity of detection and less than 35\% error in estimating the component masses for SNR > 10, thus demonstrating its ability to automatically generalize to new classes of GW signals. This is particularly promising since recent studies indicate that moderately eccentric BBH signals may be missed by quasi-circular GW searches~\cite{Huerta:2017a,Tiwari:2016,Huerta:2014,ENIGMA}. Parameter estimation on these new classes of signals is more challenging, without training, as spin and eccentricity may have the effect of mimicking features of systems of different masses. Therefore, we expect that by training on the full parameter space of signals, we will be able to decrease this error.
	
	Both our CNNs are only 23MB in size each, yet encodes all the relevant information from about 2,500 GW templates (\textasciitilde 300MB) of templates and several GB of noise used to generate the training data. The matched-filtering algorithm used for comparison required over 2s to analyze 1s inputs on our CPU. The average time taken for evaluating each of our CNNs per second of data is approximately 85 milliseconds and 540 microseconds using a single CPU and GPU respectively, thus enabling analysis even faster than real-time. While the computational cost of matched-filtering grows exponentially with the number of parameters, the Deep Filtering algorithm requires only a one-time training process, after which the analysis can be performed in constant time. Therefore, we expect the speed-up compared to matched-filtering to increase when the analysis is extended to more parameters. When considering the full range of signals that span a very high-dimensional parameter space which cannot be sampled densely due to computational costs, we expect Deep Filtering may have higher sensitivity due to its ability to interpolate and because the one-time training process can be carried out with template banks much larger than what is feasible to use with matched-filtering.

	\section{Conclusion}
	\label{conc}
	
	We showed for first time that CNNs can be used for \textit{detection} and \textit{parameter estimation} of GW signals in real LIGO data. This new paradigm for real-time GW analysis may provide the means to extend GW detection algorithms to higher dimensions and target a deeper parameter space of astrophysically motivated GW sources. The results of \texttt{Deep Filtering} can always be quickly verified via matched-filtering using a few templates near the predicted parameters. This would allow measuring additional properties such as the SNR and precise time of occurrence of the signal. Therefore, by combining \texttt{Deep Filtering} with established GW detection algorithms we may be able to enhance and accelerate GW and multimessenger campaigns and fully realize their potential for scientific discovery. 
	
	Deep learning can scale to massive training datasets~\cite{DL-Book,BigDataAI}. This would enable GW searches covering millions or billions of templates over the full range of parameter-space that is beyond the reach of existing algorithms. Extending \texttt{Deep Filtering} to predict many more parameters such as spins, eccentricities, etc., or classes of signals or glitches by adding neurons for each new parameter/class to the final layer. Furthermore, the dimensions of the CNNs can be enlarged to take inputs from multiple detectors, thus allowing coherent searches and measurements of parameters such as sky locations.
		
	Our results provide a strong incentive to extend \texttt{Deep Filtering} to cover the full range of GW signals. This study is underway, and will be described in a separate, extended article. In addition to our primary results, we have also presented several experiments exhibiting the remarkable resilience of this method for detection in periods of bad data quality, even when GW signals are contaminated with non-Gaussian transients.  This suggests that Deep Filtering could also serve as an alternative to existing glitch classification algorithms~\cite{GravitySpy,GravitySpy2,DeepTransferLearning,DeepTransferNIPS,DBNN,jade1:2016,jade:2015CQGra}. Therefore, a single, robust, and efficient data analysis pipeline for GW detectors, based on \texttt{Deep Filtering}, that unifies detection and parameter estimation along with glitch classification and clustering in real-time with very low computational overhead may potentially be built using \texttt{Deep Filtering} and deployed in the following observing runs of LIGO and Virgo in the near future.
	
	Furthermore, the initial parameters predicted by \texttt{Deep Filtering} can be used to provide instant alerts for electromagnetic follow-up campaigns and also to accelerate computationally intensive offline Bayesian parameter estimation methods~\cite{bambiann:2015PhRvD,RapidPE}. As deep CNNs excel at image processing, applying the same approach to analyze raw telescope data may accelerate the subsequent search for transient EM counterparts. Our results also suggest that, given templates of expected signals, \texttt{Deep Filtering} can be used as a generic tool for efficiently detecting and estimating properties of highly noisy time-domain signals embedded in Gaussian noise or complex non-stationary non-Gaussian noise, even in the presence of transient anomalous disturbances, in many other disciplines.

	
	\section*{Acknowledgements}
	\label{ack}
	This research is part of the Blue Waters sustained-petascale computing project, supported by the National Science Foundation (awards OCI-0725070 and ACI-1238993) and the state of Illinois. Blue Waters is a joint effort of the University of Illinois at Urbana-Champaign and its National Center for Supercomputing Applications. The eccentric numerical relativity simulations used in this article were generated with the open source, community software, the Einstein Toolkit on the Blue Waters petascale supercomputer and XSEDE (TG-PHY160053). We express our gratitude to Gabrielle Allen, Ed Seidel, Roland Haas, Miguel Holgado, Haris Markakis, Zhizhen Zhao and other members of the \href{http://gravity.ncsa.illinois.edu}{NCSA Gravity Group} along with Prannoy Mupparaju for comments and interactions and to the many others who reviewed our manuscript, including Jin Li and Kai Staats. We acknowledge the LIGO/Virgo collaboration, especially the CBC and MLA groups, for their feedback and for use of computational resources. We thank Vlad Kindratenko for granting us access to GPUs and HPC resources in the Innovative Systems Lab at NCSA. We are grateful to NVIDIA for donating several Tesla P100 GPUs and to Wolfram Research for providing technical assistance and \href{https://reference.wolfram.com}{Wolfram language} (\textit{Mathematica}) licenses which we used to carry out this research.

	\bibliographystyle{unsrt}
	\bibliography{references,references2}

\begin{thebibliography}{10}

\bibitem{LSC:2015}
{The LIGO Scientific Collaboration}, J.~{Aasi}, et~al.
\newblock {Advanced LIGO}.
\newblock {\em Classical and Quantum Gravity}, 32(7):074001, April 2015.

\bibitem{DI:2016}
B.~P. {Abbott}, R.~{Abbott}, T.~D. {Abbott}, M.~R. {Abernathy}, F.~{Acernese},
  K.~{Ackley}, C.~{Adams}, T.~{Adams}, P.~{Addesso}, R.~X. {Adhikari}, and
  et~al.
\newblock {Observation of Gravitational Waves from a Binary Black Hole Merger}.
\newblock {\em Physical Review Letters}, 116(6):061102, February 2016.

\bibitem{secondBBH:2016}
B.~P. {Abbott}, R.~{Abbott}, T.~D. {Abbott}, M.~R. {Abernathy}, F.~{Acernese},
  K.~{Ackley}, C.~{Adams}, T.~{Adams}, P.~{Addesso}, R.~X. {Adhikari}, and
  et~al.
\newblock {GW151226: Observation of Gravitational Waves from a 22-Solar-Mass
  Binary Black Hole Coalescence}.
\newblock {\em Physical Review Letters}, 116(24):241103, June 2016.

\bibitem{bbhswithligo:2016}
B.~P. {Abbott}, R.~{Abbott}, T.~D. {Abbott}, M.~R. {Abernathy}, F.~{Acernese},
  K.~{Ackley}, C.~{Adams}, T.~{Adams}, P.~{Addesso}, R.~X. {Adhikari}, and
  et~al.
\newblock {Binary Black Hole Mergers in the First Advanced LIGO Observing Run}.
\newblock {\em Physical Review X}, 6(4):041015, October 2016.

\bibitem{thirddetection}
B.~P. {Abbott}, R.~{Abbott}, T.~D. {Abbott}, M.~R. {Abernathy}, F.~{Acernese},
  K.~{Ackley}, C.~{Adams}, T.~{Adams}, P.~{Addesso}, R.~X. {Adhikari}, et~al.
\newblock {GW170104: Observation of a 50-Solar-Mass Binary Black Hole
  Coalescence at Redshift 0.2}.
\newblock {\em Physical Review Letters}, 118:221101, Jun 2017.

\bibitem{2017arXiv170909660T}
{The LIGO Scientific Collaboration}, {the Virgo Collaboration}, B.~P. {Abbott},
  R.~{Abbott}, T.~D. {Abbott}, F.~{Acernese}, K.~{Ackley}, C.~{Adams},
  T.~{Adams}, P.~{Addesso}, and et~al.
\newblock {GW170814: A Three-Detector Observation of Gravitational Waves from a
  Binary Black Hole Coalescence}.
\newblock {\em ArXiv e-prints}, September 2017.
\newblock arXiv:1709.09660 [gr-qc].

\bibitem{BNSdet:2017}
{The LIGO Scientific Collaboration} and {The Virgo Collaboration}.
\newblock {GW170817: Observation of Gravitational Waves from a Binary Neutron
  Star Inspiral}.
\newblock {\em ArXiv e-prints}, October 2017.

\bibitem{2013PhRvD..87j4028G}
D.~{Gerosa}, M.~{Kesden}, E.~{Berti}, R.~{O'Shaughnessy}, and U.~{Sperhake}.
\newblock {Resonant-plane locking and spin alignment in stellar-mass black-hole
  binaries: A diagnostic of compact-binary formation}.
\newblock {\em \prd}, 87(10):104028, May 2013.

\bibitem{Carl:2016arXiv}
C.~L. {Rodriguez}, S.~{Chatterjee}, and F.~A. {Rasio}.
\newblock {Binary black hole mergers from globular clusters: Masses, merger
  rates, and the impact of stellar evolution}.
\newblock {\em \prd}, 93(8):084029, April 2016.

\bibitem{CR:2015PRL}
C.~L. {Rodriguez}, M.~{Morscher}, B.~{Pattabiraman}, S.~{Chatterjee}, C.-J.
  {Haster}, and F.~A. {Rasio}.
\newblock {Binary Black Hole Mergers from Globular Clusters: Implications for
  Advanced LIGO}.
\newblock {\em Physical Review Letters}, 115(5):051101, July 2015.

\bibitem{Anto:2015arXiv}
F.~{Antonini}, S.~{Chatterjee}, C.~L. {Rodriguez}, M.~{Morscher},
  B.~{Pattabiraman}, V.~{Kalogera}, and F.~A. {Rasio}.
\newblock {Black Hole Mergers and Blue Stragglers from Hierarchical Triples
  Formed in Globular Clusters}.
\newblock {\em \apj}, 816:65, January 2016.

\bibitem{Samsing:2014}
J.~{Samsing}, M.~{MacLeod}, and E.~{Ramirez-Ruiz}.
\newblock {The Formation of Eccentric Compact Binary Inspirals and the Role of
  Gravitational Wave Emission in Binary-Single Stellar Encounters}.
\newblock {\em \apj}, 784:71, March 2014.

\bibitem{Huerta:2017a}
E.~A. {Huerta}, P.~{Kumar}, B.~{Agarwal}, D.~{George}, H.-Y. {Schive}, H.~P.
  {Pfeiffer}, Roland {Haas}, Wei {Ren}, T.~{Chu}, M.~{Boyle}, D.~A.
  {Hemberger}, L.~E. {Kidder}, M.~A. {Scheel}, and B.~{Szilagyi}.
\newblock {A complete waveform model for compact binaries on eccentric orbits}.
\newblock {\em \prd}, 95(2):024038, January 2017.

\bibitem{Lehner:2014a}
L.~{Lehner} and F.~{Pretorius}.
\newblock {Numerical Relativity and Astrophysics}.
\newblock {\em \araa}, 52:661--694, August 2014.

\bibitem{Sergey:2016}
S.~{Klimenko}, G.~{Vedovato}, M.~{Drago}, F.~{Salemi}, V.~{Tiwari}, G.~A.
  {Prodi}, C.~{Lazzaro}, K.~{Ackley}, S.~{Tiwari}, C.~F. {Da Silva}, and
  G.~{Mitselmakher}.
\newblock {Method for detection and reconstruction of gravitational wave
  transients with networks of advanced detectors}.
\newblock {\em \prd}, 93(4):042004, February 2016.

\bibitem{Huerta:2014}
E.~A. {Huerta}, P.~{Kumar}, S.~T. {McWilliams}, R.~{O'Shaughnessy}, and
  N.~{Yunes}.
\newblock {Accurate and efficient waveforms for compact binaries on eccentric
  orbits}.
\newblock {\em \prd}, 90(8):084016, October 2014.

\bibitem{Huerta:2013a}
E.~A. {Huerta} and D.~A. {Brown}.
\newblock {Effect of eccentricity on binary neutron star searches in advanced
  LIGO}.
\newblock {\em \prd}, 87(12):127501, June 2013.

\bibitem{2016PhRvD..94b4012H}
I.~{Harry}, S.~{Privitera}, A.~{Boh{\'e}}, and A.~{Buonanno}.
\newblock {Searching for gravitational waves from compact binaries with
  precessing spins}.
\newblock {\em \prd}, 94(2):024012, July 2016.

\bibitem{DL-Nature}
Yann Lecun, Yoshua Bengio, and Geoffrey Hinton.
\newblock Deep learning.
\newblock {\em Nature}, 521(7553):436--444, 5 2015.

\bibitem{DeepFiltering}
D.~{George} and E.~A. {Huerta}.
\newblock {Deep Neural Networks to Enable Real-time Multimessenger
  Astrophysics}.
\newblock {\em ArXiv e-prints}, December 2017.

\bibitem{lecun98-cnn}
Yann LeCun, L{\'e}on Bottou, Yoshua Bengio, and Patrick Haffner.
\newblock Gradient-based learning applied to document recognition.
\newblock {\em Proceedings of the IEEE}, 86(11):2278--2324, 1998.

\bibitem{DNNRealNoise}
D.~{George} and E.~A. {Huerta}.
\newblock {Deep Learning for Real-time Gravitational Wave Detection and
  Parameter Estimation: Results with Advanced LIGO Data}.
\newblock {\em ArXiv e-prints}, November 2017.

\bibitem{Tara:2014}
A.~{Taracchini}, A.~{Buonanno}, Y.~{Pan}, T.~{Hinderer}, M.~{Boyle}, D.~A.
  {Hemberger}, L.~E. {Kidder}, G.~{Lovelace}, A.~H. {Mrou{\'e}}, H.~P.
  {Pfeiffer}, M.~A. {Scheel}, B.~{Szil{\'a}gyi}, N.~W. {Taylor}, and
  A.~{Zenginoglu}.
\newblock {Effective-one-body model for black-hole binaries with generic mass
  ratios and spins}.
\newblock {\em \prd}, 89(6):061502, March 2014.

\bibitem{dilatedCNN}
Fisher Yu and Vladlen Koltun.
\newblock Multi-scale context aggregation by dilated convolutions.
\newblock In {\em ICLR}, 2016.

\bibitem{saton}
B.~J. {Owen} and B.~S. {Sathyaprakash}.
\newblock {Matched filtering of gravitational waves from inspiraling compact
  binaries: Computational cost and template placement}.
\newblock {\em \prd}, 60(2):022002, July 1999.

\bibitem{jade1:2016}
J.~{Powell} et~al.
\newblock {Classification methods for noise transients in advanced
  gravitational-wave detectors II: performance tests on Advanced LIGO data}.
\newblock {\em Classical and Quantum Gravity}, 34(3):034002, February 2017.

\bibitem{ETL:2012CQGra}
F.~{L{\"o}ffler}, J.~{Faber}, E.~{Bentivegna}, T.~{Bode}, P.~{Diener},
  R.~{Haas}, I.~{Hinder}, B.~C. {Mundim}, C.~D. {Ott}, E.~{Schnetter},
  G.~{Allen}, M.~{Campanelli}, and P.~{Laguna}.
\newblock {The Einstein Toolkit: a community computational infrastructure for
  relativistic astrophysics}.
\newblock {\em Classical and Quantum Gravity}, 29(11):115001, June 2012.

\bibitem{ENIGMA}
E.~A. {Huerta}, C.~J. {Moore}, P.~{Kumar}, D.~{George}, A.~J.~K. {Chua},
  R.~{Haas}, E.~{Wessel}, D.~{Johnson}, D.~{Glennon}, A.~{Rebei}, A.~M.
  {Holgado}, J.~R. {Gair}, and H.~P. {Pfeiffer}.
\newblock {ENIGMA: Eccentric, Non-spinning, Inspiral Gaussian-process Merger
  Approximant for the characterization of eccentric binary black hole mergers}.
\newblock {\em ArXiv e-prints}, November 2017.

\bibitem{chu:2016CQG}
T.~{Chu}, H.~{Fong}, P.~{Kumar}, H.~P. {Pfeiffer}, M.~{Boyle}, D.~A.
  {Hemberger}, L.~E. {Kidder}, M.~A. {Scheel}, and B.~{Szilagyi}.
\newblock {On the accuracy and precision of numerical waveforms: effect of
  waveform extraction methodology}.
\newblock {\em Classical and Quantum Gravity}, 33(16):165001, August 2016.

\bibitem{Tiwari:2016}
V.~{Tiwari}, S.~{Klimenko}, N.~{Christensen}, E.~A. {Huerta}, S.~R.~P.
  {Mohapatra}, A.~{Gopakumar}, M.~{Haney}, P.~{Ajith}, S.~T. {McWilliams},
  G.~{Vedovato}, M.~{Drago}, F.~{Salemi}, G.~A. {Prodi}, C.~{Lazzaro},
  S.~{Tiwari}, G.~{Mitselmakher}, and F.~{Da Silva}.
\newblock {Proposed search for the detection of gravitational waves from
  eccentric binary black holes}.
\newblock {\em \prd}, 93(4):043007, February 2016.

\bibitem{DL-Book}
Ian Goodfellow, Yoshua Bengio, and Aaron Courville.
\newblock {\em Deep Learning}.
\newblock MIT Press, 2016.

\bibitem{BigDataAI}
Maryam~M. Najafabadi, Flavio Villanustre, Taghi~M. Khoshgoftaar, Naeem Seliya,
  Randall Wald, and Edin Muharemagic.
\newblock Deep learning applications and challenges in big data analytics.
\newblock {\em Journal of Big Data}, 2(1):1, 2015.

\bibitem{GravitySpy}
M.~{Zevin}, S.~{Coughlin}, S.~{Bahaadini}, E.~{Besler}, N.~{Rohani},
  S.~{Allen}, M.~{Cabero}, K.~{Crowston}, A.~{Katsaggelos}, S.~{Larson}, T.~K.
  {Lee}, C.~{Lintott}, T.~{Littenberg}, A.~{Lundgren}, C.~{Oesterlund},
  J.~{Smith}, L.~{Trouille}, and V.~{Kalogera}.
\newblock {Gravity Spy: Integrating Advanced LIGO Detector Characterization,
  Machine Learning, and Citizen Science}.
\newblock {\em ArXiv e-prints}, November 2016.

\bibitem{GravitySpy2}
S.~{Bahaadini}, N.~{Rohani}, S.~{Coughlin}, M.~{Zevin}, V.~{Kalogera}, and A.~K
  {Katsaggelos}.
\newblock {Deep Multi-view Models for Glitch Classification}.
\newblock {\em ArXiv e-prints}, April 2017.

\bibitem{DeepTransferLearning}
D.~{George}, H.~{Shen}, and E.~A. {Huerta}.
\newblock {Deep Transfer Learning: A new deep learning glitch classification
  method for advanced LIGO}.
\newblock {\em ArXiv e-prints}, June 2017.

\bibitem{DeepTransferNIPS}
D.~{George}, H.~{Shen}, and E.~A. {Huerta}.
\newblock {Glitch Classification and Clustering for LIGO with Deep Transfer
  Learning}.
\newblock {\em ArXiv e-prints}, November 2017.

\bibitem{DBNN}
N.~Mukund, S.~Abraham, S.~Kandhasamy, S.~Mitra, and N.~S. Philip.
\newblock Transient classification in ligo data using difference boosting
  neural network.
\newblock {\em Phys. Rev. D}, 95:104059, May 2017.

\bibitem{jade:2015CQGra}
J.~{Powell}, D.~{Trifir{\`o}}, E.~{Cuoco}, I.~S. {Heng}, and M.~{Cavagli{\`a}}.
\newblock {Classification methods for noise transients in advanced
  gravitational-wave detectors}.
\newblock {\em Classical and Quantum Gravity}, 32(21):215012, November 2015.

\bibitem{bambiann:2015PhRvD}
J.~{Veitch}, V.~{Raymond}, B.~{Farr}, W.~{Farr}, P.~{Graff}, S.~{Vitale},
  B.~{Aylott}, K.~{Blackburn}, N.~{Christensen}, M.~{Coughlin}, W.~{Del Pozzo},
  F.~{Feroz}, J.~{Gair}, C.-J. {Haster}, V.~{Kalogera}, T.~{Littenberg},
  I.~{Mandel}, R.~{O'Shaughnessy}, M.~{Pitkin}, C.~{Rodriguez}, C.~{R{\"o}ver},
  T.~{Sidery}, R.~{Smith}, M.~{Van Der Sluys}, A.~{Vecchio}, W.~{Vousden}, and
  L.~{Wade}.
\newblock {Parameter estimation for compact binaries with ground-based
  gravitational-wave observations using the LALInference software library}.
\newblock {\em \prd}, 91(4):042003, February 2015.

\bibitem{RapidPE}
C.~{Pankow}, P.~{Brady}, E.~{Ochsner}, and R.~{O'Shaughnessy}.
\newblock {Novel scheme for rapid parallel parameter estimation of
  gravitational waves from compact binary coalescences}.
\newblock {\em \prd}, 92(2):023002, July 2015.

\end{thebibliography}

\end{document}